\documentclass[aps,prl,twocolumn,showpacs,superscriptaddress]{revtex4}
\usepackage{graphicx}
\begin{document}

% Use the \preprint command to place your local institutional report
% number in the upper righthand corner of the title page in preprint mode.
% Multiple \preprint commands are allowed.
% Use the 'preprintnumbers' class option to override journal defaults
% to display numbers if necessary
%\preprint{}

%Title of paper
\title{$\mu$SR studies of RE(O,F)FeAs (RE = La, Nd, Ce) and LaOFeP systems:  
possible incommensurate/stripe magnetism and superfluid density}
% repeat the \author .. \affiliation  etc. as needed
% \email, \thanks, \homepage, \altaffiliation all apply to the current
% author. Explanatory text should go in the []'s, actual e-mail
% address or url should go in the {}'s for \email and \homepage.
% Please use the appropriate macro foreach each type of information
% \affiliation command applies to all authors since the last
% \affiliation command. The \affiliation command should follow the
% other information
% \affiliation can be followed by \email, \homepage, \thanks as well.
     \author{J.~P.~Carlo} 
     \affiliation{Department of Physics, Columbia University, New York, New York 10027, USA}
     \author{Y.~J.~Uemura}
     \altaffiliation[author to whom correspondences should be addressed: E-mail
tomo@lorentz.phys.columbia.edu]{}
     \affiliation{Department of Physics, Columbia University, New York, New York 10027, USA}
    \author{T.~Goko}
     \affiliation{Department of Physics, Columbia University, New York, New York 10027, USA}
     \affiliation{Department of Physics and Astronomy, McMaster University, Hamilton, Ontario L8S 4M1, Canada}
     \affiliation{TRIUMF, 4004 Wesbrook Mall, Vancouver, B.C., V6T 2A3, Canada} 
     \author{G.~J.~MacDougall}
     \author{J.~A.~Rodriguez}
     \author{W.~Yu}
     \author{G.~M.~Luke}
     \affiliation{Department of Physics and Astronomy, McMaster University, Hamilton, Ontario L8S 4M1, Canada}
     \author{Pengcheng~Dai}
     \affiliation{Department of Physics and Astronomy, University of Tennessee, Knoxville, Tennessee 37996, USA}
     \author{N. Shannon}
     \affiliation{H H Wills Physics Laboratory, University of Bristol, BS8 1TL Bristol, United Kingdom}
     \author{S.~Miyasaka}
     \author{S.~Suzuki}
     \author{S.~Tajima}
     \affiliation{Department of Physics, Osaka University, 1-1 Machikaneyama, Toyonaka, Osaka 560-0043, Japan}
     \author{G.~F.~Chen}
     \author{W.~Z.~Hu}
     \author{J.~L.~Luo}
     \author{N.~L.~Wang}
     \affiliation{Beijing National Laboratory for Condensed Matter Physics, Institute of Physics,
Chinese Academy of Sciences, Beijing 100080, Peoples Republic of China}

\date{\today}

\begin{abstract}
Muon spin relaxation ($\mu$SR) measurements in iron oxy-pnictide 
systems have revealed: (1) commensurate long-range order in undoped LaOFeAs;
(2) Bessel function line shape in  
La(O$_{0.97}$F$_{0.03}$)FeAs which indicates possible incommensurate or stripe
magnetism; (3) anomalous weak magnetism existing in superconducting
LaOFeP, Ce(O$_{0.84}$F$_{0.16}$)FeAs, and Nd(O$_{0.88}$F$_{0.12}$)FeAs but
absent in superconducting La(O$_{0.92}$F$_{0.08}$)FeAs;
and (4) scaling of superfluid density and $T_{c}$ in the Ce, La, and Nd-FeAs superconductors following a nearly
linear relationship found in cuprates.
\end{abstract}

% insert suggested keywords - APS authors don't need to do this
%\keywords{}
%\maketitle must follow title, authors, abstract, \pacs, and \keywords
\maketitle

%para 1

A renewed interest on high-$T_{c}$ superconductivity has been
generated by recent discoveries of iron-oxypnictide superconductors LaOFeP 
($T_{c} \sim$ 5K) \cite{laofephosono} and LaOFeAs ($T_{c} \sim$ 26K) \cite{laofeashosono}, 
followed by subsequent development of materials with higer $T_{c}$'s
up to $\sim$ 55K containing rare earth (RE) 
elements, such as Ce, Nd, Sm \cite{xhchen,gfchen,zren} instead of La.   Carriers are doped by (O,F) or 
(La,Sr) \cite{holedopedhhwen}
substitutions as well as high-pressure oxygen 
synthesis \cite{highpressureoxygen}.  For overall understanding
of mechanisms of high-$T_{c}$ superconductivity, it would be
very intstructive to compare these new
superconductors with cuprate systems.  Muon spin 
relaxation ($\mu$SR) studies have provided unique information
on magnetic order \cite{uemuralscoprl,brewerprl,saviciprb}
and superfluid density 
\cite{uemuraprl89,uemuraprl91,reviewrmp,reviewscot}.
in cuprate systems.
In this letter, we report $\mu$SR measurements on undoped and doped 
iron-oxypnictide systems, LaOFeP, LaOFeAs,
La(O$_{0.97}$F$_{0.03}$)FeAs, La(O$_{0.92}$F$_{0.08}$)FeAs,
Ce(O$_{0.84}$F$_{0.16}$)FeAs and Nd(O$_{0.88}$F$_{0.12}$)FeAs.
Our results demonstrate several generic features common to
the iron-oxypnictides and cuprate systems, including long-range
commensurate antiferromagnetism of the undoped parenet compounds,
incommensurate or stripe magnetism of lightly doped systems
near the border of magnetic and superconducting phases, and
scaling of the superfluid density and $T_{c}$ following a
nearly common linear relationship.  We also report
anomalous weak magnetism detected in some of the superconducting 
systems.

%para2

Polycrystalline specimens of all the FeAs compounds were synthesized at
Beijing National Laboratory for Condensed Matter Physics,
following the method described in \cite{nlwangmethod}.  
Specimen of LaOFeP was synthesized at Osaka University,
in the silica tubes 
using LaP, Fe and Fe$_{2}$O$_{3}$ as starting materials.
These specimens were made
into ceramic pellets of $\sim$ 8 mm in diameter and 1 $\sim$ 2 mm
thick.  The specimens of undoped LaOFeAs and superconducting
La(O$_{0.92}$F$_{0.08}$)FeAs are identical to those used in neutron scattering
measurements \cite{daineutron} which revealed collinear antiferromagnetic
order of LaOFeAs below $\sim$ 134 K and absence of long range magnetism
in the superconducting La(O$_{0.92}$F$_{0.08}$)FeAs.
$\mu$SR measurements were performed at TRIUMF using a He gas-flow
cryostat in zero field (ZF), longitudinal field (LF), and
weak-transverse-field (WTF) to characterize magnetism, 
and in transverse field (TF) to determine
the magnetic field penetration depth.  Details of the $\mu$SR
method can be found in refs. \cite{reviewrmp,reviewscot}.

%para3

Figure 1(a) shows the time spectra of muon spin polarization
in zero field observed in the undoped parent system LaOFeAs. 
The long-lived and very clear muon spin 
precession signal indicates spatially long-ranged and homogenous
magnetism, consistent with the commensurate Bragg peak 
for collinear antiferromagnetic structure found by neutron studies \cite{daineutron,lumsdenneutron}.
The $\mu$SR spectra can be fitted to a high frequency $f(T\rightarrow 0) \sim 23$ MHz
signal with a dominant ($\sim$ 60 \%\ volume fraction) signal amplitude, and additional signal with 
much lower frequency $f(T \rightarrow 0) \sim$ 3 MHz from a 
minor ($\sim$ 10\%\ ) volume fraction.  The temperature
dependences of these frequencies are shown in Fig. 2(a). 
These results in LaOFeAs are essentially consistent with earlier reports by
$\mu$SR and Moessbaurer studies \cite{klausscondmat,japanesemoessbauer} in LaOFeAs.
Comparison of sub-lattice magnetization ($\propto f(T)$) with spin-wave theories will
be published separately.

\begin{figure}[t]
\includegraphics[width=3.2in,angle=0]{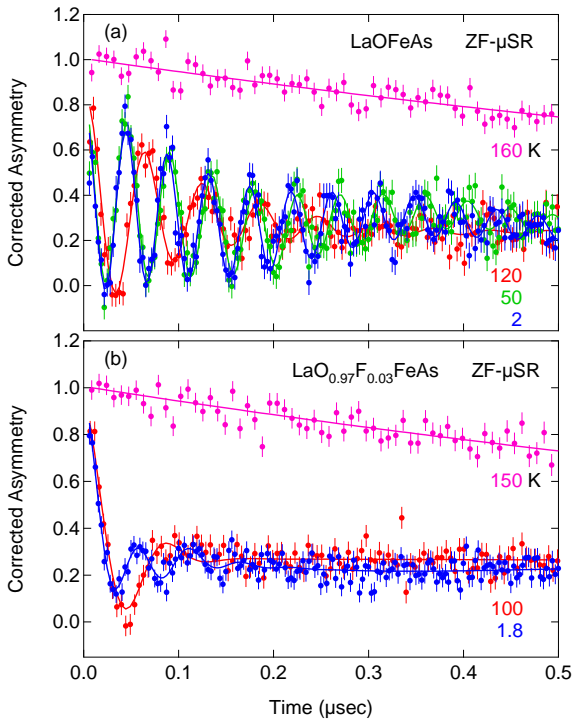}%
\caption{\label{fig1}(color)
Time spectra of Zero-Field $\mu$SR in (a) undoped LaOFeAs
and (b) 3\%\-doped La(O$_{0.97}$F$_{0.03}$)FeAs.
The solid lines in (b) for T = 100 K represents a single
Bessel function multiplied to a Gaussian damping,
while for T = 2 K a sum of three Bessel*Gaussian signals.}
\end{figure}

\begin{figure}[h]
\includegraphics[width=3.2in,angle=0]{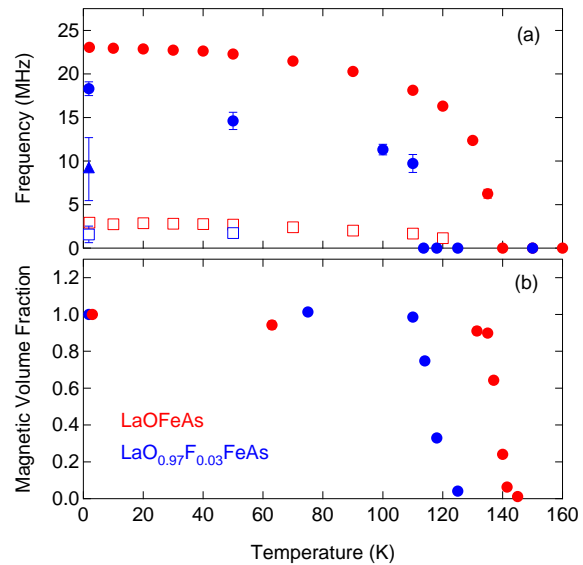}%
\caption{\label{fig2}(color)
(a) Muon spin precession frequency observed in zero field
in LaOFeAs (for two cosine signals) and La(O$_{0.97}$F$_{0.03}$)FeAs
(for three Bessel signals).
(b) The magnetic volume fraction estimated by WTF measurements
with the external field of 50 G.}
\end{figure}

%para4

ZF-$\mu$SR spectra of the 3\%\ doped La(O$_{0.93}$F$_{0.07}$)FeAs, shown in Fig. 1(b),
exhibit much faster damping than those in undoped LaOFeAs.  The fast-damping
spectra in Fig. 1(b) fit well to Bessel functions (multiplied by a Gaussian damping factor),
which are shown by the solid lines.  
The Bessel function line shape was first observed in ZF-$\mu$SR  
in an incommensurate spin density wave (SDW) system TMTSF$_{2}$PF$_{6}$ \cite{lpletmtsfprb}.  Subsequently,
this line-shape was also observed by ZF-$\mu$SR in La$_{1.875}$Ba$_{0.125}$CuO$_{4}$ \cite{lukelbcozf},     
La$_{1.47}$Nd$_{0.4}$Sr$_{0.13}$CuO$_{4}$ and 
several other cuprate systems \cite{saviciprb} which undergo formation of static
spin stripes.  The comparison of parent and doped systems in Fig. 1 exhibits exactly the same 
features as in the case of the cuprates, La$_{2}$CuO$_{4}$ and the 1/8 doped spin stripe systems, 
shown in Fig. 4(b), 2(a) and 2(b) of ref. \cite{saviciprb}.
On the other hand, it is in principle possible to expect a highly damped spectra 
also in a commensurate antiferromagnet with substantial randomness, such as electron-doped
cuprate systems in the antiferromagnetic region \cite{lukenature,lukeprb}.
Neutron scattering studies are required for a clear distinction of these three spin structures 
and determination of the spatial periodicity.

%para 5

Figure 2(a) compares the ZF precession frequencies of the two cosine signals in undoped
LaOFeAs (closed and open red symbols) with the frequencies of the Bessel function signals
in the 3\%\ -doped compound (closed blue symbol).  In the case of incommensurate SDW order, 
the frequency of the Bessel function 
corresponds to the internal field at the muon site near the maximum SDW amplitude.  
The lower Bessel frequencies at T = 2 K in the 3\%\-doped system implies that the 
average internal field at the muon site is substantially reduced from that 
in the undoped LaOFeAs.  This is remiscent to the case of 
cuprates, where the ZF-$\mu$SR frequency in La$_{2}$CuO$_{4}$ was about $\sim$ 30 \%\ higher
than the Bessel frequency in the 1/8 doped stripe systems, as shown in Fig. 3(b) of ref. \cite{saviciprb}.
Such a large variation of internal field via 3\%\ carrier doping is hardly expected by
a simple magnetic dilution of commensurate antiferromagnets.  
Although $\mu$SR is a local magnetic probe which cannot determine precise spin structures,
these considerations suggest a strong possibility of incommensurate or static stripe magnetism
in La(O$_{0.93}$F$_{0.07}$)FeAs. 

%para 6
Figure 2(b) shows the volume fraction
of magnetically ordered region in the undoped and doped FeAs systems, obtained in the 
WTF-$\mu$SR measurements \cite{uemuramnsi113} with WTF $\sim$ 50 G, after a correction for a background
signal with the spectral weight of less than 10\%\ of the total signal amplitude.
In both doped and undoped systems, the magnetic order develops in essentially full volume 
fraction below the Neel temperture $T_{N}$.  A continuous change with finite volume fraction was observed
in a narrow range of temperature within +/- 5 K of $T_{N}$ in both systems, indicating a weakly
first order nature of thermal phase transition.
A rather high $T_{N} \sim$ 110 K and full magnetic volume fraction 
of the doped system indicate robustness of static magnetism against carrier doping,
which is reminiscent to the case of non-superconducting electron-doped cuprates. 
%Figure 2(c) shows the low temperature ZF-$\mu$SR frequency in LaOFeAs, which is proportional to the 
%sub-lattice magnetization $M_{sub}$.  The variation of $M_{sub}$  
%reflects thermal excitation of spin waves near the antiferromagnetic zone center.
%The results for the undoped system, fitting well to the $T^{?}$ power below T = ? K, 
%indicate: (1) little Ising anisotropy in the exchange interaction; and (2) existence of relatively 
%soft spin-wave branch presumably due to interlayer coupling.  

\begin{figure}[t]
\includegraphics[width=3.5in,angle=0]{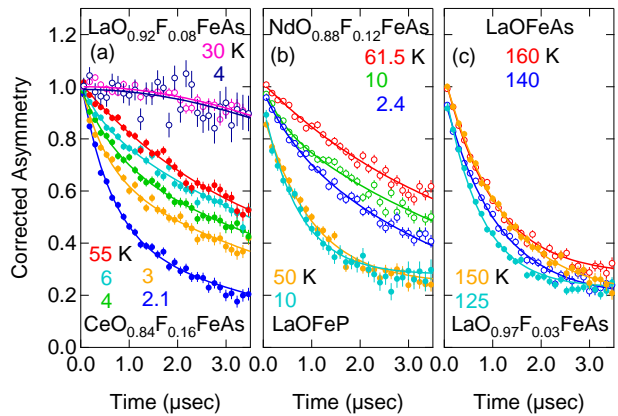}%
\caption{\label{fig3}(color)
Muon spin relaxation function observed in zero field in
(a) La and Ce-FeAs superconductors, (b) Nd-FeAs and LaOFeP
superconductors, and (c) in the parmagnetic state of 
LaOFeAs and La(O$_{0.97}$F$_{0.03}$)FeAs.} 
%The temperature dependences in the Ce-FeAs system below T = 6K
%in (a),
%and in the Nd-FeAs system at T = 2.4 K in (b), are presumably due to imminent magnetic order of Ce and Nd moments.}
\end{figure}

 %para7 

ZF-$\mu$SR measurements have also revealed an anomalous and weak
relaxation existing in normal and superconducting
states of  LaOFeP,
Ce(O$_{0.84}$F$_{0.16}$)FeAs and Nd(O$_{0.88}$F$_{0.12}$)FeAs,
as well as in the paramagnetic state of LaOFeAs and
La(O$_{0.97}$F$_{0.03}$)FeAs above $T_{N}$.  As shown by the time spectra in Fig. 3(a)-(c),
this relaxation is seen in most of the systems in the present study, except for  
the superconducting La(O$_{0.92}$F$_{0.08}$)FeAs.
The anomalous relaxation has also been reported in SmO$_{0.82}$F$_{0.18}$FeAs
\cite{isissmcondmat}, while the absence of the effect in the La-based superconducting
system is consistent with an earlier report \cite{luetkenscondmat}.  
These ZF-$\mu$SR spectra exhibit almost no temperature dependence, except for 
the Ce compound below $T \sim$ 4 K and Nd compound at T = 2.4 K, which is presumably
due to imminent ordering of Ce or Nd moments.  The static origin of this
anomalous relaxation was confirmed by LF-$\mu$SR measuremens in LaOFeP at T = 8 and 2K,
while static and additional dynamic effects were found for 
the Sm compound in ref. \cite{isissmcondmat} at T = 60 K.
The observed relaxation rate of 0.3 - 1 $\mu s^{-1}$ corresponds to about 3 - 10 G
of random static internal field.  If it comes from dilute frozen moments of $\sim$ 1 Bohr
magneton, this field corresponds to 0.01 - 0.1 \%\ level of concentration per formula unit. 
These features indicate: (1) the slow anomalous relaxation does not necessarily
correlate with superconducting transition; (2) this effect exists not only in systems
containing magnetic rare earth elements but also in several La based systems;
(3) dilute frozen moments, conceivable for a non-stoichiometry
of Fe and/or rare-earth sites and/or for minority impurity phases, 
are possible origin of the observed effect, but
further studies are required for conclusive determination.  

%para8
\begin{figure}[t]
\includegraphics[width=3.5in,angle=0]{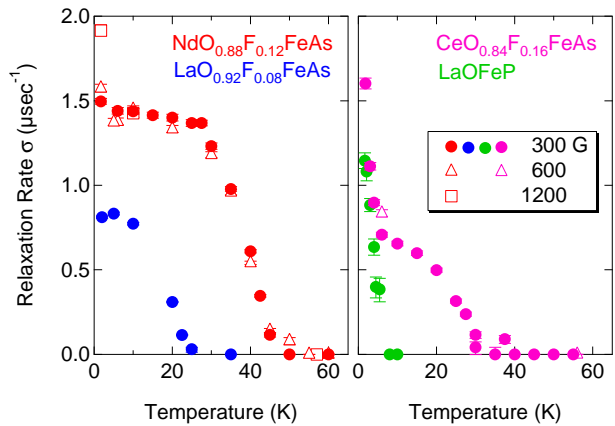}%
\caption{\label{fig4}(color)
Muon spin relaxation rate $\sigma$ observed in transverse external
fields in superconducting FeAs and FeP based systems.
The relaxation envelope was fit to a pre-fixed exponential
function multiplied by a Gaussian function
$exp(-\sigma^{2}t^{2}/2)$.  Absence of field dependence
is consistent with the effect due to the magnetic field penetration 
depth.}
\end{figure}

On the four superconducting specimens (those in Figs. 3(a)and (b)), we
performed TF-$\mu$SR measurements in TF = 300, 600, and 1200 G to 
measure the magnetic field penetration depth $\lambda$, and to derive the 
superfluid density $n_{s}/m^{*}$ (superconducting carrier density / effective mass) 
proportional to the relaxation rate $\sigma$ due to flux vortex 
lattice below $T_{c}$.  To correct for the effect of the anomalous
magnetic relaxation shown in Fig. 3, we first obtained an exponential relaxation 
rate in TF above $T_{c}$, fixed this value in the fitting, multipled a Gaussian
precession envelope $exp(-\sigma^{2}t^{2}/2)$ to this exponential damping,
and derived $\sigma$ as shown in Fig. 4. 
This procedure does not make essential difference from a simpler fitting with 
Gaussian damping alone in the FeAs based superconductors, since the pre-fixed exponential 
relaxation rate (0.03, 0.24 and 0.34 $\mu s^{-1}$, respectively, for the La, Nd, and Ce compound)
is much smaller than the superconducting Gaussian relaxation rate.
For LaOFeP, the exponential (1.36 $\mu s^{-1}$) and Gaussian relaxation rates
are very close, resulting in a significant systematic uncertainty in 
accuracy of the superfluid density.     
The rapid increase of $\sigma$ in the Nd compound (at T=2K) and Ce compound (below T = 4 K)
are presumably due to imminent magnetic order of moments on these elements.
Otherwise, $\sigma(T)$ shows a saturation at low temperatures, consistent with an earlier
report \cite{luetkenscondmat}.  In view of difficulty in determining
the pairing symmetry using $\mu$SR results on ceramic specimens experienced in the cuprates, however,
we postpone the symmetry arguments until results on single-crystal specimens become
available.

%para9 

In the $\sigma(T\rightarrow 0)$ vs. $T_{c}$ plot of Fig. 5, we compare the results of 
the FeAs based systems to cuprates and a few other exotic superconductors. 
Three points from the present study and two other points from ref. \cite{luetkenscondmat}
demonstrate that the electron-doped FeAs systems follow a nearly linear relationship between
$T_{c}$ and $n_{s}/m^{*} \propto \sigma(T\rightarrow 0)$, with the slope approximately same as that
found for cuprates.  We note that our earlier results on an electron-doped cuprate
(Nd,Ce)$_{2}$CuO$_{4}$ \cite{lukeelectrondope} also follow the behavior of hole-doped cuprates.
%and the results in ref. \cite{electrondopeprl} might involve some effect of magnetism \cite{kojimaelectrondope}.
%Further studies are required to determine whether electron-doped cuprates follow the trend of 
%hole-doped cuprates in this plot.
Figure 5 deomnstrates that many type-II superconductors, including cuprates, iron-oxypnictides,
and $C_{60}$ systems, share approximately same ratios between their superfluid energy scales and $T_{c}$.
This implies that $n_{s}/m^{*}$ is an important factor in determining $T_{c}$.

\begin{figure}[t]
\includegraphics[width=3.2in,angle=0]{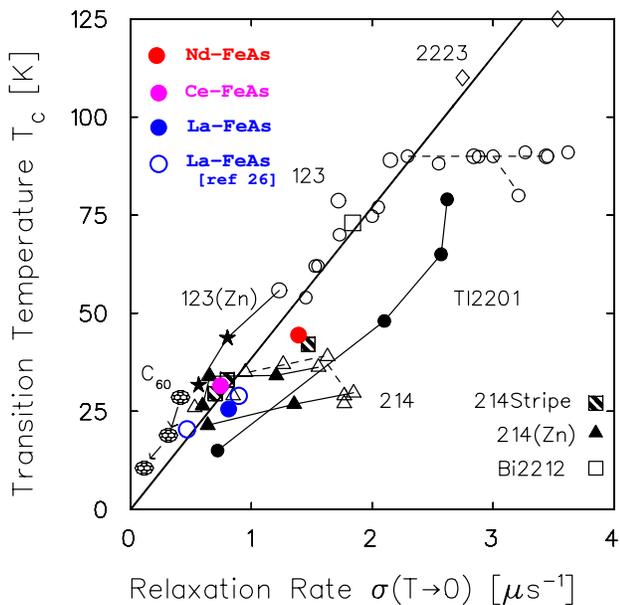}%
\caption{\label{fig5}(color)
A plot of the muon spin relaxation rate $\sigma(T\rightarrow 0)
\propto n_{s}/m^{*}$ versus $T_{c}$ for the 
iron oxypnictide systems (colored solid circles for the present
results and colored open circles for ref. \cite{luetkenscondmat} compared with the results for
the cuprates and alkali-doped C$_{60}$ systems (\cite{uemuraprl89,uemurajpcm,yamazakiprize}).
The FeAs superconductors follow the linear relationship found for underdoped cuprates in this plot.}
\end{figure}

%para10  
A large relaxation rate $\sigma(T\rightarrow 0)$ observed in a rather low $T_{c}$ LaOFeP, however,
suggests that additional factor(s), such as closeness to the magnetically ordered state,
also plays a significant role in determining $T_{c}$, as discussed in ref. \cite{uemurajpcm,yamazakiprize} for cuprates.
In these references, one of us \cite{uemurajpcm,yamazakiprize} proposed a paring mechanism
for cuprates based on a charge motion (with an effective Fermi energy $kT_{F}$ derived from the superfluid density)
resonanting with antiferromagnetic spin flucturations (with an energy scale $\hbar\omega_{AF}$
represented by the exchange interaction $J$)
as $kT_{F} \sim \hbar\omega_{AF}$.  
With the present results suggesting $\sigma(T\rightarrow 0) \propto n_{s}/m^{*} \propto kT_{F}$ (in 2-d sysems) 
$\propto T_{c}$, this resonant
spin-charge motion model might also be applicable to oxy-pnictides if their $\hbar\omega_{AF}$
is approximately 50 meV or so, as estimated in a theoretical work \cite{yildirim}.
In this model, superconductivity should be absent in the doping region with static Cu/Fe magnetism
(except for phase separation), which is the case both in the copper oxide and FeAs superconductors.  
Inelastic neutron and two-magnon Raman studies of $J$ and $\hbar\omega_{AF}$ of FeAs systems would be quite
interesting from this view point.
In summary, we demonstrated several common features between cuprates and oxy-pnictides
by comparing $\mu$SR results on their magnetism and superfluid density.

We acknowledge financial support from
US NSF DMR-05-02706 (Material World Network) at Columbia,
NSERC and CIAR (Canada) at McMaster, Grant-In-Aid for Scientific Reasearch (No. 19014012,
19204038) from the MEXT (Japan) at Osaka U., and NSFC, CAS, and 973 project of MOST
(China) at IOP, Beijing.
\\

\vfill \eject

\begin{thebibliography}{99}
%\bibitem[*]{byline} author to whom correspondences should be addressed. email: 
%tomo@lorentz.phys.columbia.edu

\bibitem{laofephosono} 
Y. Kamihara {\it et al.\/}, J. Am. Chem. Soc. {\bf 128\/} (2006) 10012.
\bibitem{laofeashosono}
Y. Kamihara, T. Watanabe, M. Hirano, H. Hosono,
J. Am. Chem. Soc. {\bf 130\/} (2008) 3296.
\bibitem{xhchen}
X.H. Chen {\it et al.\/},
Cond-mat: arXiv:0803.3603 (2008)
\bibitem{gfchen}
G.F. Chen {\it et al.\/}, Cond-mat arXiv:0803.3790 (2008).
\bibitem{zren}
Zhi-An Ren {\it et al.\/}., Cond-mat arXiv:0803.4283 (2008). 
\bibitem{holedopedhhwen} 
 H.H. Wen, G. Mu, L. Fang, H. Yang, X. Zhu,
Europhys. Lett. {\bf 82\/} (2008) 17009.
\bibitem{highpressureoxygen}
Zhi-An Ren {\it et al.\/}, Cond-mat arXiv:0804.2582 (2008)
\bibitem{uemuralscoprl}
Y.J. Uemura {\it et al.\/}, Phys. Rev. Lett. {\bf 59\/} (1987) 1045.
\bibitem{brewerprl}
J.H. Brewer {\it et al.\/}, Phys. Rev. Lett. {\bf 60\/} (1988) 1073.
\bibitem{saviciprb}
A.T. Savici {\it et al.\/}, Phys. Rev. {\bf B66\/} (2002)014524.
\bibitem{uemuraprl89}
Y.J. Uemura {\it et al.\/}, Phys. Rev. Lett. {\bf 62\/} (1989) 2317.
\bibitem{uemuraprl91}
Y.J. Uemura {\it et al.\/}, Phys. Rev. Lett. {\bf 66\/} (1991) 2665.
\bibitem{reviewrmp}
J.E. Sonier, J.H. Brewer, R.F. Kiefl, 
%{\sl \ \ \ $\mu$SR Studies of the Vortex State in Type-II Superconductors\/},
Rev. Mod. Phys. {\bf 72\/} (2002) 769. 
\bibitem{reviewscot} {\it Muon Science: Muons in Physics, Chemistry and
Materials\/}, ed. by S.L. Lee, S.H. Kilcoyne, and R. Cywinski,
Inst. of Physics Publishing, Bristol, 1999.
\bibitem{nlwangmethod}
J. Dong {\it et al.\/}, Cond-mat arXiv:0803.3426 (2008)
\bibitem{daineutron}
C. de la Cruz {\it et al.\/}, Cond-mat arXiv:0804.0795 (2008), to appear in Nature.
\bibitem{lumsdenneutron}
M.B. Stone {\it et al.\/}, Cond-mat arXiv:0801.2332 (2008)
\bibitem{klausscondmat}
H.H. Klauss {\it et al.\/},
Cond-mat arXiv:0805.0264 (2008)
\bibitem{japanesemoessbauer}
S. Kitao {\it et al.\/},
Cond-mat arXiv:0805.0041 (2008).
\bibitem{lpletmtsfprb}
L.P. Le {\it et al.\/},
Phys. Rev. {\bf B48\/} (1993) 7284.
\bibitem{lukelbcozf}
G.M. Luke {\it et al.\/},
Physica {\bf C185\/}-{\bf 189\/} (1991) 1175.
\bibitem{lukenature}
G.M. Luke {\it et al.\/}, Nature {\bf 338\/} (1989) 49. 
\bibitem{lukeprb}
G.M. Luke {\it et al.\/}, Phys. Rev. {\bf B42\/} (1990) 7981.
\bibitem{uemuramnsi113}
Y.J. Uemura {\it et al.\/}, Nature Physics {\bf 3\/} (2007) 29.
\bibitem{isissmcondmat} 
A.J. Drew {\it et al.\/}, Cond-mat arXiv:0805.1042 (2008)
\bibitem{luetkenscondmat}
H. Luetkens {\it et al.\/},
Cond-mat arXiv:0804.3115 (2008)
\bibitem{lukeelectrondope}
G.M. Luke {\it et al.\/}, 
Physica {\bf C282\/}-{\bf 287\/} (1997) 1465.
\bibitem{uemurajpcm}
Y.J. Uemura, J. Phys. Condens. Matter {\bf 16\/} (2004) S4515.
\bibitem{yamazakiprize}
Y.J. Uemura,
Physica {\bf B374\/}-{\bf 375\/} (2006) 1.
\bibitem{yildirim}
T. Yildirim, Cond-mat arXiv:0804.2252 (2008). 
\end{thebibliography}
\end{document}